\documentclass[amsmath,amssymb,prc,final,showpacs,twocolumn]{revtex4-1}
 
\usepackage{bm,hyphenat,xspace}
\usepackage{graphicx,epsfig}

\newcommand {\mbf}[1]{{\mathbf{#1}}}

\newcommand {\mcu}{\mathcal{U}}

\newcommand{\cm}{\mathrm{c\!\:\!.m\!\:\!.}}
\newcommand{\He}{{}^3\mathrm{He}}
\newcommand{\Hh}{{}^3\mathrm{H}}
\newcommand{\nH}{n\text{-}{}^3\mathrm{H}}
\newcommand{\pHe}{p\text{-}{}^3\mathrm{He}}

\begin{document}

\title {Calculation of proton-${}^3$He elastic scattering between 7 and 35 MeV}
 
\author{A.~Deltuva} 
\affiliation{Centro de F\'{\i}sica Nuclear da Universidade de Lisboa, 
P-1649-003 Lisboa, Portugal }

\author{A.~C.~Fonseca} 
\affiliation{Centro de F\'{\i}sica Nuclear da Universidade de Lisboa, 
P-1649-003 Lisboa, Portugal }

\received{March 21, 2013}
\pacs{21.45.-v, 25.10.+s, 21.30.-x, 24.70.+s}

\begin{abstract}
\begin{description}
\item[Background]
Theoretical calculations of the four-particle scattering above
the four-cluster breakup threshold are technically very difficult due to
nontrivial singularities or boundary conditions.
Further complications arise when the long-range Coulomb force is present.
\item[Purpose]
We aim at calculating proton-${}^3$He elastic scattering observables
above three- and four-cluster breakup threshold.
\item[Methods]
We employ Alt, Grassberger, and Sandhas (AGS) equations for the 
four-nucleon transition operators and solve them in the momentum-space
framework using the complex-energy method whose accuracy and practical 
applicability is improved by a special integration method.
\item[Results]
Using realistic nuclear interaction models we obtain fully converged results
for the proton-${}^3$He elastic scattering. The differential cross section,
proton and ${}^3$He analyzing powers, spin correlation and spin transfer
coefficients  are calculated at proton energies ranging from
7 to 35 MeV. Effective three- and four-nucleon forces are included via 
the explicit excitation of a nucleon to a $\Delta$ isobar.
\item[Conclusions]
Realistic  proton-${}^3$He scattering calculations above the four-nucleon 
breakup threshold are feasible. There is quite good agreement between the 
theoretical predictions and experimental data for the proton-${}^3$He 
scattering in the considered energy regime. The most remarkable
disagreements are the peak of the proton analyzing power at 
lower energies and the minimum of the differential cross section
at higher energies. Inclusion of the $\Delta$ isobar reduces the latter
discrepancy.
\end{description}
\end{abstract}

 \maketitle

\section{Introduction \label{sec:intro}}

Proton-${}^3$He ($\pHe$) scattering is one of the most commonly used 
experiments to study the four-nucleon system \cite{tilley:92a}:
It involves two charged particles that are stable and easy to 
detect with acceptable precision, there are no competing channels 
until $E_p= 7.3$ MeV proton lab energy,  and beyond that 
only three- and four-cluster breakup takes place up to the pion-production
threshold.
Much like neutron-${}^3$H ($\nH$) scattering, $\pHe$ is dominated by 
isospin $T=1$, but has the deciding experimental advantage of 
having a proton beam and a non-radioactive ${}^3$He target.
 On the contrary, from the 
theoretical point of view,  $\pHe$ is more difficult to calculate 
than $\nH$ due to the long-range Coulomb force between protons ($p$)
that gives rise to complicated boundary conditions 
in the coordinate-space and non-compact kernel in the momentum-space
representation.
Nevertheless, these difficulties have been solved below the
 three-cluster breakup threshold  using 
three different theoretical frameworks, 
namely, the hyperspherical harmonics (HH) expansion method
\cite{viviani:01a,kievsky:08a}, the Faddeev-Yakubovsky (FY) equations 
\cite{yakubovsky:67}  for the wave function components in coordinate space
\cite{lazauskas:04a,lazauskas:09a}, and
the Alt, Grassberger and Sandhas (AGS) equations 
\cite{grassberger:67} for transition operators
that were solved in the momentum space \cite{deltuva:07a,deltuva:07b}.
A good agreement between these methods has been  
demonstrated in a benchmark 
for $\nH$ and $\pHe$ elastic scattering observables \cite{viviani:11a}
using realistic nucleon-nucleon (NN) potentials.

Recently we extended the AGS calculations to energies
above three- and four-cluster breakup thresholds
\cite{deltuva:12c,deltuva:13a}. 
The complex energy method \cite{kamada:03a,uzu:03a} was used to deal with 
the complicated singularities in the four-particle scattering equations;
its accuracy and practical applicability
was greatly improved by a special integration method
 \cite{deltuva:12c}. This allowed us  to achieve fully
converged results for  $\nH$ elastic scattering and
 neutron-neutron-deuteron recombination into $n+\Hh$
using realistic NN interactions.
We note that the FY calculations
of $\nH$ elastic scattering have been recently extended as well
to energies above the four-nucleon breakup threshold
\cite{lazauskas:12a}, however, using a semi-realistic
NN potential  limited to $S$-waves.

In the present work we extend the method of Ref.~\cite{deltuva:12c}  
to calculate the   $\pHe$ elastic scattering 
above breakup threshold and compare with existing data for cross sections 
and spin  observables over a wide range of proton beam energies up to 
$E_p=35$ MeV. The $pp$ Coulomb interaction 
is treated as in Refs.~\cite{deltuva:05a,deltuva:07b} using the 
method of screening of the $pp$ Coulomb potential 
followed by the phase renormalization of transition amplitudes
\cite{taylor:74a,alt:80a}.
Thus, standard AGS scattering equations with short-range potentials
are applicable. At energies bellow three-cluster
threshold our results agree with those obtained by other methods as 
mentioned in  Ref.~\cite{viviani:11a}. Compared to our previous
$\nH$ scattering calculations above the breakup threshold \cite{deltuva:12c},
the most serious complication for $\pHe$ 
is the  convergence of the partial-wave expansion
that requires a larger number of states due to the longer range
of the screened Coulomb potential.

In Sec.~\ref{sec:eq} we describe the theoretical formalism
and in Sec.~\ref{sec:res} we present the numerical results.
The summary is given in  Sec.~\ref{sec:sum}.

\section{4N scattering equations \label{sec:eq}}

We  use the symmetrized AGS equations \cite{deltuva:07a}
as appropriate for the four-nucleon system in the isospin formalism.
They are integral equations for the four-particle transition operators 
$\mcu_{\beta \alpha}$, i.e.,
\begin{subequations}  \label{eq:AGS}   
\begin{align}  
\mcu_{11}  = {}&  -(G_0 \, t \, G_0)^{-1}  P_{34} -
P_{34} U_1 G_0 \, t \, G_0 \, \mcu_{11}  \nonumber \\ 
{}& + U_2   G_0 \, t \, G_0 \, \mcu_{21}, \label{eq:U11}  \\
\label{eq:U21}
\mcu_{21} = {}&  (G_0 \, t \, G_0)^{-1}  (1 - P_{34})
+ (1 - P_{34}) U_1 G_0 \, t \, G_0 \, \mcu_{11}.
\end{align}
\end{subequations}
Here, $\alpha=1$ corresponds to the $3+1$ partition (12,3)4
whereas  $\alpha=2$ corresponds to the $2+2$ partition (12)(34);
there are no other distinct  two-cluster partitions in the system
of four identical particles.
The free resolvent at the complex energy $E+ i\varepsilon$ is given by
\begin{gather}\label{eq:G0}
G_0 = (E+ i\varepsilon - H_0)^{-1},
\end{gather} 
with  $H_0$ being the free Hamiltonian.
The pair (12) transition matrix
\begin{gather} \label{eq:t}
t = v + v G_0 t
\end{gather} 
is derived from the potential $v$;
for the $pp$ pair  $v$ includes both the nuclear and the 
screened Coulomb potential $w_R$. Our calculations are done
in momentum space; however, we start with the configuration-space
representation 
\begin{equation} \label{eq:wr}
w_R(r) = w(r) \, e^{-(r/R)^n},
\end{equation}
where  $w(r) = \alpha_e/r$ is the true Coulomb potential, 
$\alpha_e \simeq 1/137$ is the fine structure constant, 
$R$ is the screening radius,
and $n$ controls the smoothness of the screening.
All transition operators acquire parametric dependence on $R$
but it is suppressed in our notation, except for the 
scattering amplitudes.
The symmetrized 3+1 or 2+2 subsystem transition operators are
obtained from the respective integral equations
\begin{gather} \label{eq:AGSsub}
U_\alpha =  P_\alpha G_0^{-1} + P_\alpha t\, G_0 \, U_\alpha.
\end{gather}
The basis states are antisymmetric under exchange of two particles in the 
subsystem (12) for the $3+1$ partition 
and in (12) and (34) for the $2+2$ partition.
The full antisymmetry of the four-nucleon system is ensured by the 
permutation operators $P_{ab}$ of particles $a$ and $b$ with
$P_1 =  P_{12}\, P_{23} + P_{13}\, P_{23}$ and $P_2 =  P_{13}\, P_{24}$.

The $\pHe$ scattering amplitude with nuclear plus screened Coulomb
interactions
at available energy $E = \epsilon_1 + 2p_1^2/3m_N$
 is obtained from the on-shell matrix element 
$  \langle \mbf{p}_{1}'| T_{(R)} |\mbf{p}_{1} \rangle
  = 3 \langle  \phi_{1}' | \mcu_{11}| \phi_{1} \rangle $
in the limit $\varepsilon \to +0$. Here 
$|\phi_{1} \rangle $ is the Faddeev component of the
asymptotic  $\pHe$ state in the channel $\alpha=1$, characterized
by the bound state energy $\epsilon_1 = -7.72$ MeV and
the relative $\pHe$ momentum $\mbf{p}_1$, 
$m_N$ being the average nucleon mass.
Due to energy conservation  $p_1' = p_1$.

The amplitude $  \langle \mbf{p}_{1}'| T_{(R)} |\mbf{p}_{1} \rangle$
is decomposed into its
long-range part $\langle \mbf{p}_1'| t_R^{\cm} |\mbf{p}_1 \rangle$, 
being the two-body on-shell transition matrix derived 
from the screened Coulomb potential of the form \eqref{eq:wr} between the 
proton and the center of mass (c.m.)  of $\He$, 
and the remaining Coulomb-distorted short-range part.
Renormalizing $  \langle \mbf{p}_{1}'| T_{(R)} |\mbf{p}_{1} \rangle$
by the phase factor $Z_R^{-1}$ \cite{taylor:74a,alt:80a,deltuva:07b},
in the $R \to \infty$ limit, yields the full $\pHe$ transition 
amplitude in the presence of Coulomb
\begin{gather}      \label{eq:T} 
  \begin{split}
    \langle \mbf{p}_1'| T |\mbf{p}_1 \rangle  = {} &
    \langle \mbf{p}_1'| t_C^{\cm} |\mbf{p}_1 \rangle   \\
    &+ \lim_{R \to \infty} \left\{ 
    \langle \mbf{p}_1'| [ T_{(R)} - t_R^{\cm} ]|\mbf{p}_1 \rangle
    Z_R^{-1}  \right\},
    \end{split}
\end{gather}  
where  the first term is obtained from
 $Z_R^{-1} \langle \mbf{p}_1'| t_R^{\cm} |\mbf{p}_1 \rangle$  that converges,  
in general as a distribution, to the 
exact Coulomb amplitude $\langle \mbf{p}_1'| t_C^{\cm} |\mbf{p}_1 \rangle$ 
between the proton and the c.m. of the $\He$ nucleus \cite{taylor:74a}.
The renormalization factor $Z_R$ is defined in 
Refs.~\cite{deltuva:05a,deltuva:07b}.
The second term in Eq.~\eqref{eq:T}, 
after renormalization by $Z_R^{-1}$, represents the Coulomb-modified
nuclear short-range amplitude. It has to be calculated numerically,
but, due to its short-range nature, the $R \to \infty$ limit
is reached with sufficient accuracy at finite screening radii $R$.
Since the convergence with $R$ is faster at higher energies,
the required screening radii are smaller than in our low-energy 
$\pHe$ calculations \cite{deltuva:07b}. 
We found that $R$ ranging from 8 to 10 fm 
leads to well-converged results in the energy regime considered in the
present paper. Furthermore, we take a sharper screening 
with $n=6$  such that at short distances
$r < R$ the screened Coulomb approximates  the full Coulomb better
than with $n=4$ used in Ref.~\cite{deltuva:07b}
and at the same time vanishes more rapidly at $r > R$
thereby accelerating the partial-wave convergence.

We solve the AGS equations \eqref{eq:AGS} in the momentum-space
partial-wave framework. 
The states of the  total angular momentum  $\mathcal{J}$ 
with the projection  $\mathcal{M}$ are defined as  
$ | k_x \, k_y \, k_z   
[l_z (\{l_y [(l_x S_x)j_x \, s_y]S_y \} J_y s_z ) S_z] \,\mathcal{JM} \rangle$ 
for the $3+1$ configuration and 
$|k_x \, k_y \, k_z  (l_z  \{ (l_x S_x)j_x\, [l_y (s_y s_z)S_y] j_y \} S_z)
\mathcal{ J M} \rangle $ for the $2+2$.
Here  $k_x , \, k_y$ and $k_z$ are the four-particle Jacobi momenta
in the convention of Ref.~\cite{deltuva:12a}, 
$l_x$, $l_y$, and $l_z$ are the associated orbital angular momenta,
$j_x$ and $j_y$ are the total angular momenta of pairs (12) and (34),
$J_y$ is the total angular momentum of the (123) subsystem,
 $s_y$ and $s_z$ are the spins of nucleons 3 and 4, 
and $S_x$, $S_y$, and $S_z$ are channel spins
of two-, three-, and four-particle system.
A similar coupling scheme is used for the isospin.
We include a large number of four-nucleon partial 
waves, up to  $l_x,l_y,l_z,j_x,j_y = 7$,  $J_y = \frac{13}{2}$,
 and $\mathcal{J} = 6$,
such that the results are well converged.
In fact, lower cutoffs are sufficient for lower $\mathcal{J}$, e.g.,
$l_x,l_y,l_z,j_x,j_y \le 5$ and  $J_y \le \frac{9}{2}$ are sufficient
for $\mathcal{J} \le 3$. Furthermore, for most observables 
 $\mathcal{J} \le 5$ or even  $\mathcal{J} \le 4$ are enough
for the convergence; $\mathcal{J} = 6$ yields small but still visible
effect only above $E_p=30$ MeV.

The numerical calculations are performed for complex energies, i.e.,
with finite  $\varepsilon$. The limit 
 $\varepsilon \to +0$ needed for the calculation of the amplitude
$  \langle \mbf{p}_{1}'| T_{(R)} |\mbf{p}_{1} \rangle$ is obtained by the
extrapolation of finite  $\varepsilon$ results as proposed
in Ref.~\cite{kamada:03a}. A special integration method
developed in  Ref.~\cite{deltuva:12c} is used 
to treat the quasi-singularities of the AGS equations \eqref{eq:AGS}.
We obtain accurate results by using $ \varepsilon$ ranging
from 1 to 2 MeV at lowest considered energies and from
2 to 4 MeV at highest considered energies. Grid points for the 
discretization of each momentum variable range from 30 (at lower energies)
to 35 (at higher energies). Further
details on the other numerical techniques for solving the four-nucleon 
AGS equations can be found in Ref.~\cite{deltuva:07a}.

\section{Results \label{sec:res}}

We study the $\pHe$ scattering  using  realistic high-precision
NN potentials, namely, the Argonne (AV18) potential  \cite{wiringa:95a},
the inside-nonlocal outside-Yukawa
(INOY04) potential  by Doleschall \cite{doleschall:04a,lazauskas:04a},
the charge-dependent Bonn potential (CD Bonn)  \cite{machleidt:01a},
and its extension CD Bonn + $\Delta$ \cite{deltuva:03c}
allowing for an excitation of a nucleon to a $\Delta$ isobar
and thereby yielding effective three- and four-nucleon forces
(3NF and 4NF).
The $\He$ binding energy  calculated with AV18, CD Bonn, CD Bonn + $\Delta$,
and INOY04 potentials is 6.92, 7.26, 7.54, and 7.73 MeV, respectively;
the experimental value is 7.72 MeV.
Therefore most of our predictions correspond to  INOY04 as it is the
only potential that nearly reproduces the experimental binding energy of 
$\He$. The calculations with other potentials are done at fewer selected
energies.

\begin{figure}[!]
\begin{center}
\includegraphics[scale=0.66]{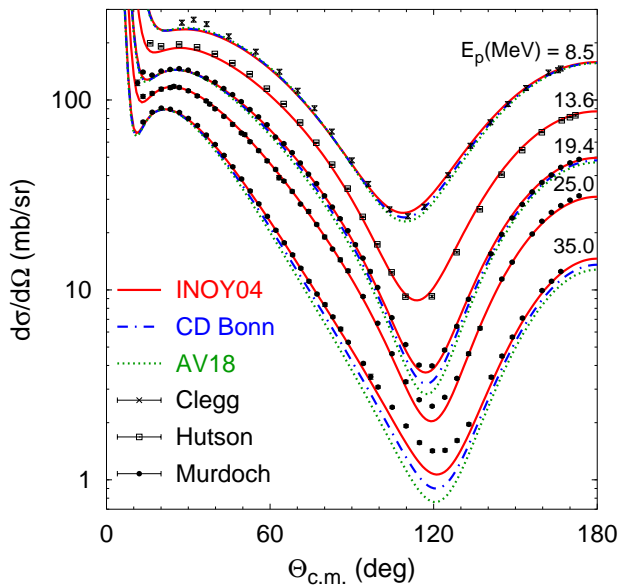}
\end{center} 
\caption{ \label{fig:dcs1} (Color online) 
Differential cross section for elastic $\pHe$ scattering at 
8.52, 13.6, 19.4, 25.0, and 35.0 MeV proton energy
as function of the c.m. scattering angle.
Results obtained with INOY04 (solid curves), and, at selected energies,
with CD Bonn (dashed-dotted curves) and AV18 (dotted curves) 
potentials are compared with the experimental data from
Refs.~\cite{clegg:64,hutson:71a,murdoch:84a}.}
\end{figure}

\begin{figure}[!]
\begin{center}
\includegraphics[scale=0.66]{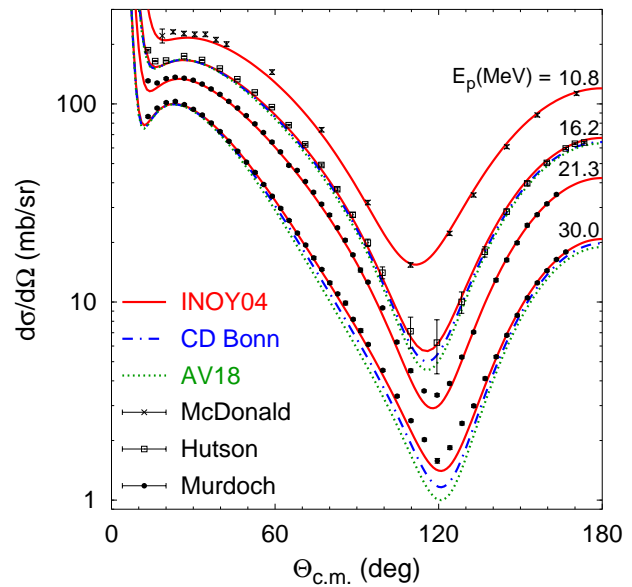}
\end{center} 
\caption{ \label{fig:dcs2} (Color online) 
Same as in Fig.~\ref{fig:dcs1} but at
10.77, 16.23, 21.3, and 30.0 MeV proton energy.
The experimental data are from
Refs.~\cite{mcdonald:64,hutson:71a,murdoch:84a}.}
\end{figure}

In Figs.~\ref{fig:dcs1} - \ref{fig:dcs2} we show the 
differential cross section $d\sigma/d\Omega$ for elastic $\pHe$ scattering
as a function of the c.m. scattering angle $\Theta_{\cm}$
at a number of proton energies ranging from $E_p = 8.5$ to 35.0 MeV.
This observable decreases rapidly with the increasing energy 
and also changes the shape; the calculations
describe the energy and angular dependence of the experimental data
fairly well.
Below $E_p = 15$ MeV the  experimental data are slightly underpredicted at
forward angles as happens also at energies below the three-cluster breakup
threshold \cite{viviani:11a,deltuva:07b}.
At  the minimum the $d\sigma/d\Omega$ predictions scale
with the $\He$ binding energy:  the weaker the   
$\He$  binding the lower the dip of $d\sigma/d\Omega$ that is located 
between $\Theta_{\cm} = 105^{\circ}$ and $\Theta_{\cm} = 125^{\circ}$.
The scaling is more pronounced at higher  $E_p$.
 For the INOY04 potential that fits the $\He$ binding energy, 
one gets an good  agreement in the whole angular region up to 
$E_p\simeq 20$ MeV but, as the energy increases, the calculated cross section 
underpredicts the data at the minimum much like what happens in 
nucleon-deuteron elastic scattering \cite{witala:98a,nemoto:98c,deltuva:05a} 
but for nucleon energies above  60 MeV. 
In line with the conjectures that were made 15 years ago 
for the three-nucleon system \cite{witala:98a,nemoto:98c}, 
this underprediction of the data at the minimum of $d\sigma/d\Omega$
may be a sign for the need to include the 3NF. 

\begin{figure*}[!]
\begin{center}
\includegraphics[scale=0.8]{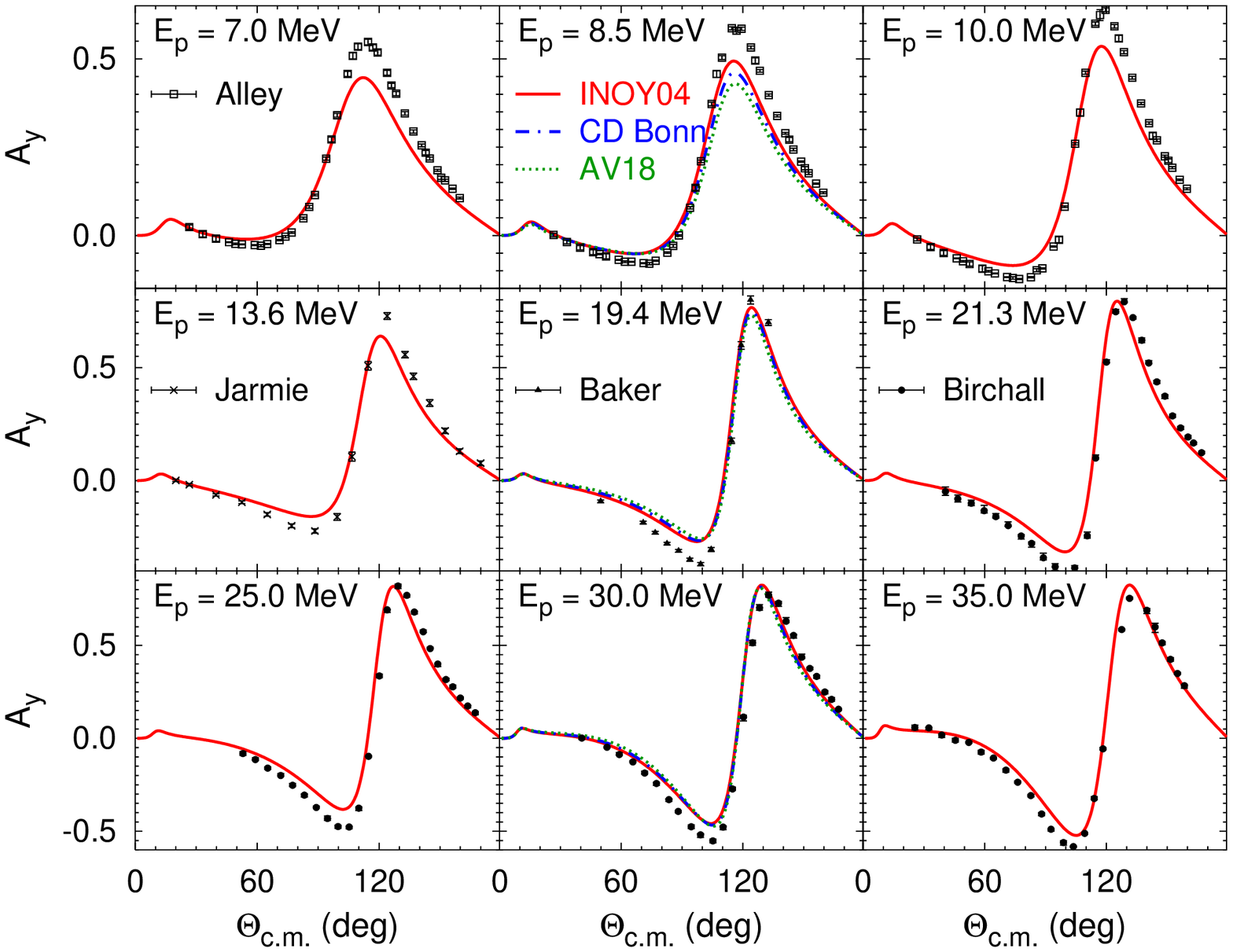}
\end{center} 
\caption{ \label{fig:ay} (Color online) 
Proton analyzing power  $A_{y}$ for elastic $\pHe$ scattering
at 7.03, 8.52, 10.03, 13.6, 19.4, 21.3, 25.0, 30.0, and 35.0 MeV
proton energy. Curves as in Fig.~\ref{fig:dcs1}.
The experimental data are from
Refs.~\cite{alley:93,jarmie:74a,baker:71a,birchall:84a}.}
\end{figure*}

In Fig.~\ref{fig:ay}  we show the proton analyzing power  $A_{y}$
for elastic $\pHe$ scattering
at proton energies ranging from 7.0 to 35.0 MeV. We observe that the 
sensitivity to the nuclear force model  and
energy  is considerably weaker as compared to the regime
below three-cluster threshold \cite{deltuva:07a,deltuva:07b}.
Most remarkably,
in contrast to low energies where the famous $p$-$\He$ $A_y$-puzzle
exists \cite{viviani:01a,fisher:06,deltuva:07b},  the peak of $A_y$
around 120 degrees is described fairly well but
there is a discrepancy in the minimum region.
This is similar to the energy evolution of the  $A_y$-puzzle in the
three-nucleon system  \cite{gloeckle:96a}.

\begin{figure*}[!]
\begin{center}
\includegraphics[scale=0.8]{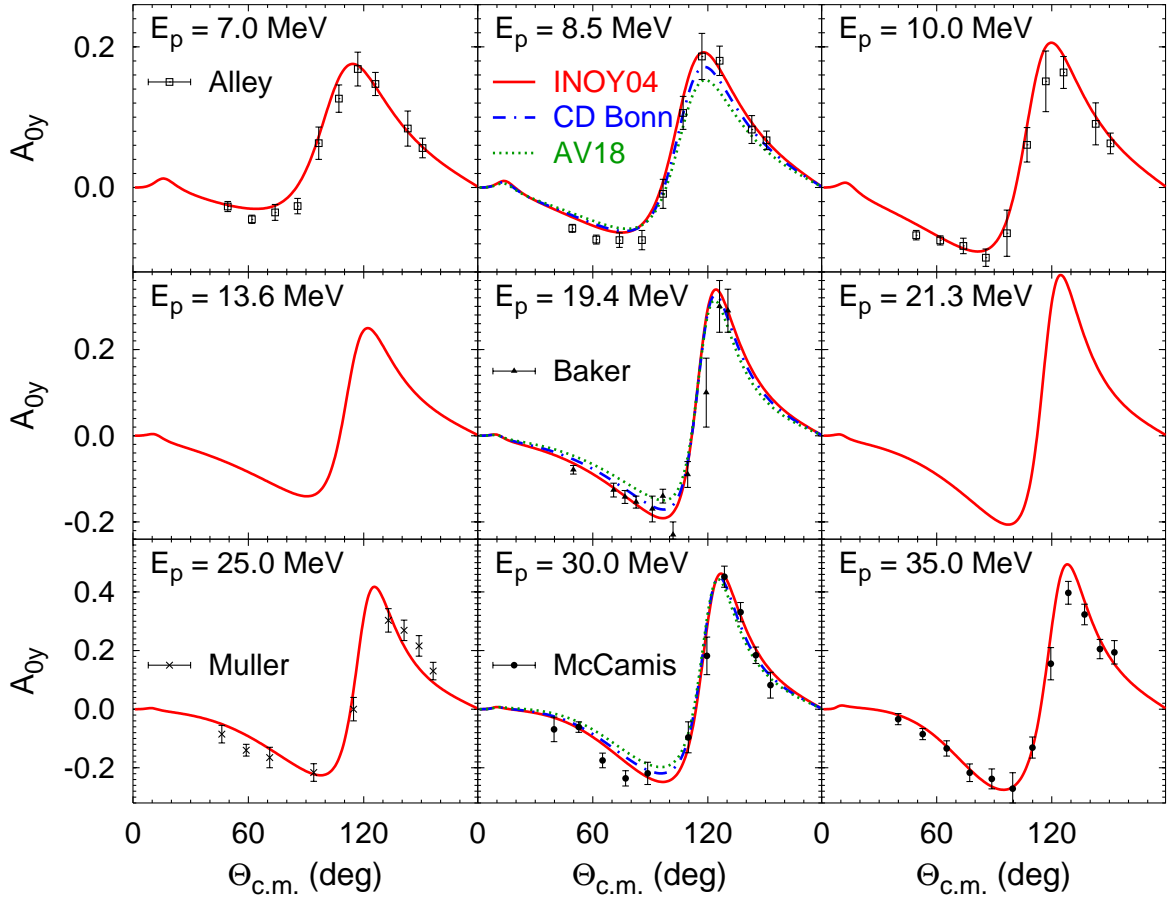}
\end{center} 
\caption{ \label{fig:a0y} (Color online) 
$\He$ analyzing power $A_{0y}$ for elastic $\pHe$ scattering
at 7.03, 8.52, 10.03, 13.6, 19.4, 21.3, 25.0, 30.0, and 35.0 MeV
proton energy. Curves as in Fig.~\ref{fig:dcs1}.
The experimental data are from
Refs.~\cite{alley:93,baker:71a,muller:78a,mccamis:85a}.}
\end{figure*}

In Fig.~\ref{fig:a0y}  we show the 
$\He$ analyzing power  $A_{0y}$
for elastic $\pHe$ scattering
at proton energies ranging from 7.0 to 35.0 MeV. 
 $A_{0y}$ varies slowly with energy but is slightly more sensitive to the
NN potential. Contrary to  $A_y$, 
calculated $A_{0y}$ is in better agreement with data over the whole energy 
range, particularly when the INOY04 potential is used.

\begin{figure}[!]
\begin{center}
\includegraphics[scale=0.63]{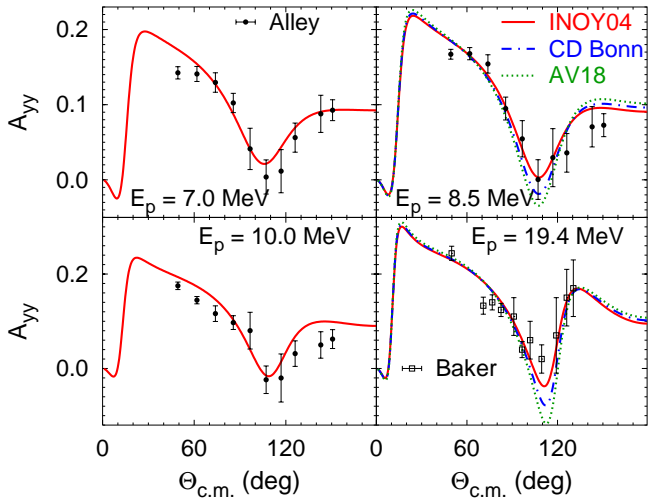}
\end{center} 
\caption{ \label{fig:ayy} (Color online) 
$\pHe$  spin correlation coefficient $A_{yy}$
 for elastic $\pHe$ scattering
at 7.03, 8.52, 10.03, and 19.4 MeV proton energy. 
Curves as in Fig.~\ref{fig:dcs1}.
The experimental data are from Refs.~\cite{alley:93,baker:71a}.}
\end{figure}

\begin{figure}[!]
\begin{center}
\includegraphics[scale=0.67]{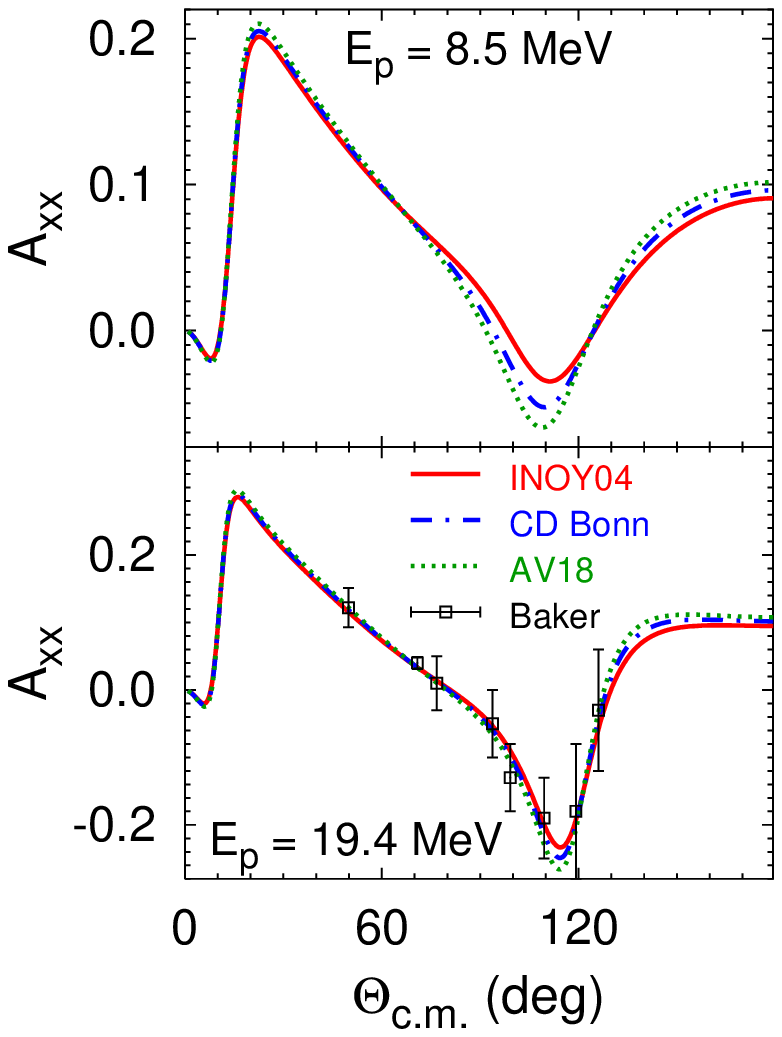}
\end{center} 
\caption{ \label{fig:axx} (Color online) 
$\pHe$  spin correlation coefficient $A_{xx}$
 for elastic $\pHe$ scattering
at  8.52 and 19.4 MeV proton energy. 
Curves as in Fig.~\ref{fig:dcs1}.
The experimental data are from Ref.~\cite{baker:71a}.}
\end{figure}

The experimental data are scarcer for double polarization observables.
In Fig.~\ref{fig:ayy}  we show the 
$\pHe$  spin correlation coefficient $A_{yy}$
for elastic $\pHe$ scattering
at 7.03, 8.52, 10.03, and 19.4 MeV, and in Fig.~\ref{fig:axx}  we show $A_{xx}$
for elastic $\pHe$ scattering
at 8.52 and 19.4 MeV proton energy. Calculated $A_{yy}$ 
exhibits some sensitivity to the NN potential model and
describes the data  reasonably well; 
the agreement with data is the best when the INOY04
interaction is used. The same happens for $A_{xx}$ but for the single 
data set we know of.

\begin{figure*}[!]
\begin{center}
\includegraphics[scale=0.8]{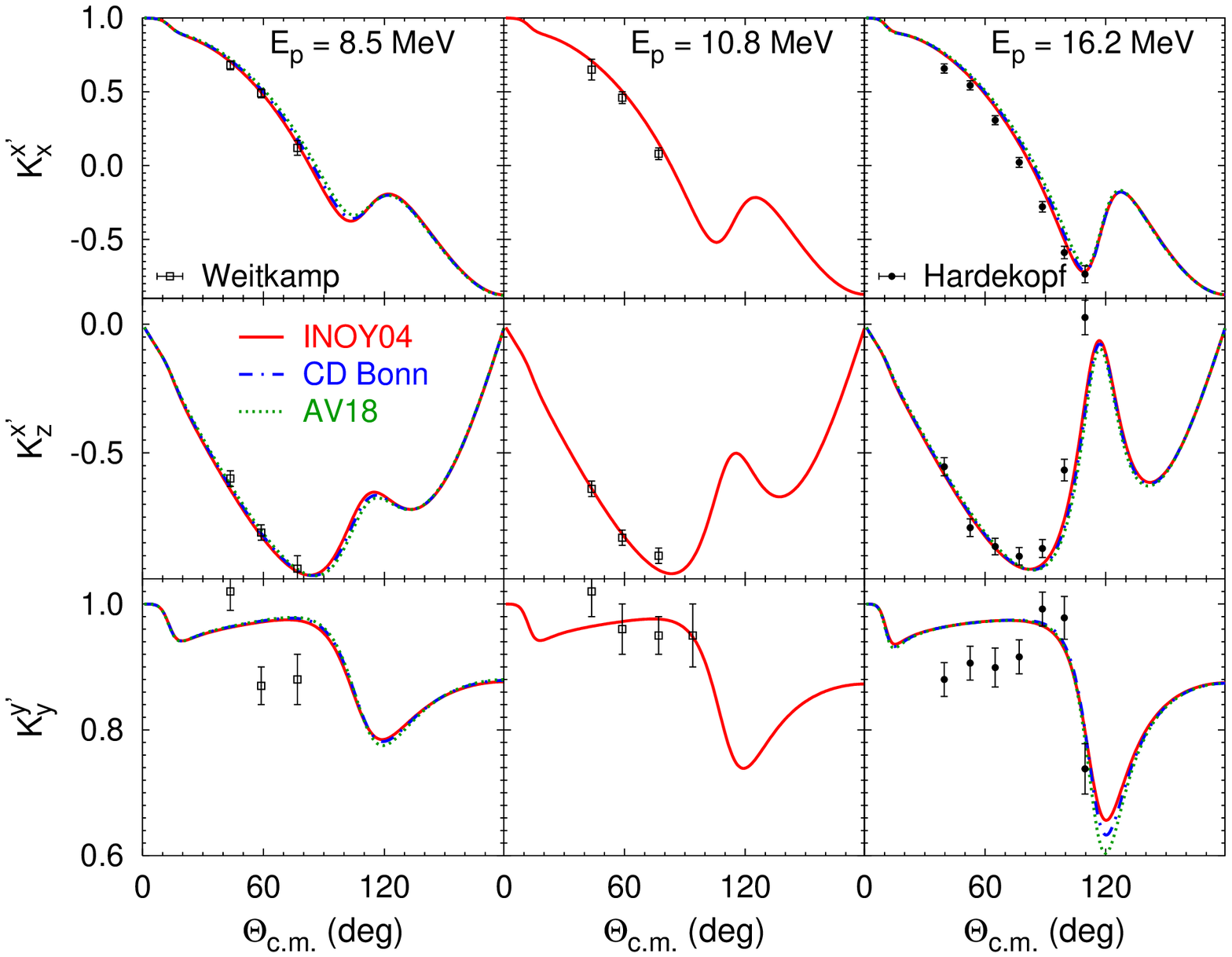}
\end{center} 
\caption{ \label{fig:k} (Color online) 
Proton spin transfer coefficients $K_{x}^{x'}$, $K_{z}^{x'}$, and $K_{y}^{y'}$
for elastic $\pHe$ scattering
at 8.52, 10.77, and 16.23 MeV proton energy. 
Curves as in Fig.~\ref{fig:dcs1}.
The experimental data are from
Refs.~\cite{weitkamp:78a,hardekopf:76a}.}
\end{figure*}

Finally in Fig.~\ref{fig:k}  we show the 
proton spin transfer coefficients $K_{x}^{x'}$, $K_{z}^{x'}$, and $K_{y}^{y'}$
for elastic $\pHe$ scattering at 8.52, 10.77, and 16.23 MeV proton energy. 
Note that 8.52-MeV predictions are compared to experimental data taken
at 8.82 MeV but, given the weak energy dependence of these observables,
the comparison  is appropriate.
The calculated spin transfer coefficients show a rich angular structure 
and follow the data reasonably well but
cannot be fully tested by the available data 
confined to the angular region below  $\Theta_{\cm} = 110^{\circ}$.
In contrast to other shown spin observables, the 
spin transfer coefficient $K_{z}^{x'}$ around $\Theta_{\cm} = 120^{\circ}$,
i.e., in the region of the differential cross section minimum,
varies quite rapidly with the energy.
There is little sensitivity to NN interaction model, except for
$K_{y}^{y'}$ around $\Theta_{\cm} = 120^{\circ}$ at $E_p=16.23$ MeV.
Theoretical results and experimental data for $K_{y}^{y'}$ are close
to 1 up to $\Theta_{\cm} = 90^{\circ}$, but data sets at different energies
seem to be inconsistent  as they show different angular dependence.
In contrast, theoretical predictions at the three considered energies
show nearly the same angular dependence for $\Theta_{\cm} \le 90^{\circ}$.

\begin{figure}[!]
\begin{center}
\includegraphics[scale=0.6]{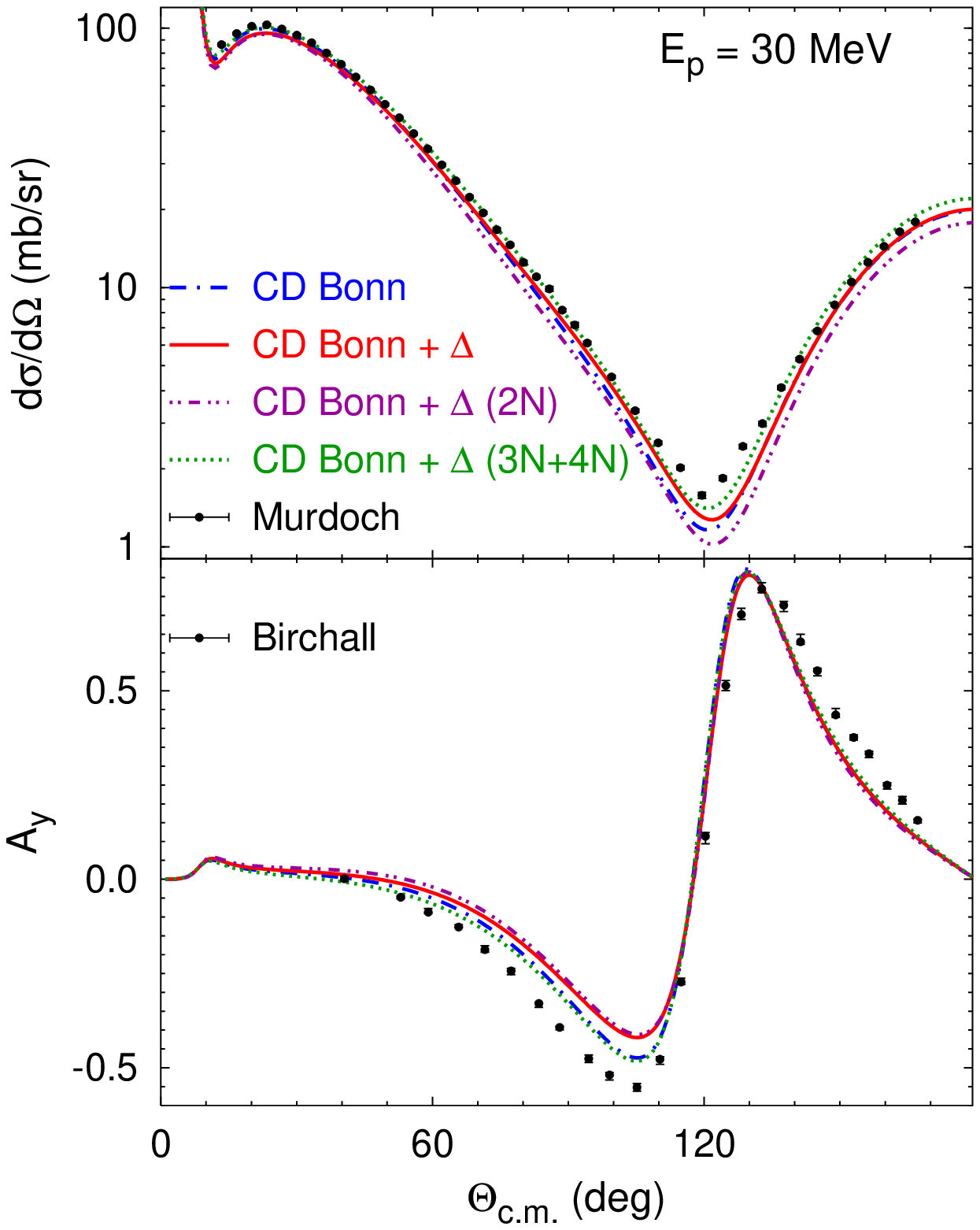}
\end{center} 
\caption{ \label{fig:delta} (Color online) 
Differential cross section and proton analyzing power  $A_{y}$
for elastic $\pHe$ scattering at 30 MeV proton energy. 
Results obtained with CD Bonn (dashed-dotted curves) and 
CD Bonn + $\Delta$ (solid curves) potentials
are compared with the experimental data from
Refs.~\cite{murdoch:84a,birchall:84a}. In addition are shown 
the predictions including $\Delta$ effects of 2N nature
(dashed-double-dotted curves) and of 3N and 4N nature (dotted curves).}
\end{figure}

As already mentioned, above $E_p=20$ MeV the minimum 
of the elastic differential cross section is underpredicted. 
In order to establish the importance of the 3NF as a means 
to cure this discrepancy we study the effect of the $\Delta$-isobar
excitation on both the differential cross section and proton analyzing power  
at 30 MeV proton energy. This has been done before at energies below 
three-cluster breakup threshold \cite{deltuva:08a} and we follow here the 
same procedure. The results are shown in 
Fig.~\ref{fig:delta} in a way that one can single out the 
$\Delta$-isobar effects of 2N nature, the so-called 2N dispersion, 
and of 3N and 4N nature, the  3NF and 4NF. The competition between 
 2N dispersion and 3NF,
often found in the 3N system,  is well seen also here for the 
differential cross section. As shown in Fig.~\ref{fig:delta} 
dispersive effect increases the discrepancy with data 
(dashed-double dotted curves) while  3NF and 4NF effects reverse that trend 
for the differential cross section (dotted curves). 
Nevertheless, when the two effects are put together 
the net result is an improvement towards the data (solid curves) 
but not quite enough to bridge the original gap.
For $A_y$ only the dispersive effect around the minimum
is visible; it moves the predictions away from data.

\section{Summary \label{sec:sum}}

In summary, we performed fully converged proton-$\He$ elastic
scattering calculations with realistic potentials
above the three- and four-cluster breakup threshold. 
The symmetrized Alt, Grassberger,
and Sandhas four-particle equations were solved in the momentum-space
framework.
We  used the complex energy method whose
accuracy and the efficiency is greatly improved
by the numerical integration technique with special weights.
The $pp$ Coulomb interaction was included rigorously using the
method of screening and renormalization.

The differential cross section exhibits rapid energy dependence
and, in the minimum region around $\Theta_{\cm} = 120^{\circ}$,
also sensitivity to the NN interaction model.
The calculations using the INOY04 potential 
describe the experimental data well up to $E_p = 20$ MeV but
underpredict the differential cross section in the minimum at higher energies;
other potential models fail even more.
In contrast, most of the calculated spin observables show little sensitivity
to the interaction model, and also the dependence on the beam
energy is weaker than below the three-cluster breakup threshold.
The overall agreement with the 
experimental data for the spin observables is quite good, 
considerably better than in the low-energy 
 $p$-$\He$ scattering which is  affected by $P$-wave resonances. 
In particular, the peak of the proton analyzing power $A_y$ that is
strongly underpredicted at low energies, is reproduced fairly well
above $E_p = 20$ MeV but there is discrepancy in the minimum.
The observed sensitivity to the  NN interaction model seems
to be mostly due to different predictions of the $\He$ binding energy;
the calculations using the INOY04 potential with correct $\He$ binding
provide the best description of the experimental data.
 
We also studied the effect of three- and four-nucleon forces 
through the explicit inclusion of the $\Delta$-isobar excitation.
We found that $\Delta$-generated many-nucleon forces significantly
improve the description of the differential cross section  but have almost
no effect for the proton analyzing power. However, there are
also quite strong dispersive  $\Delta$-isobar effects that often 
reduce or even reverse the effect of 3NF. Therefore
the total $\Delta$-isobar effect, although beneficial, is not large enough
to bridge the gap between the differential cross section data and calculations.
It even increases the discrepancy in the minimum of $A_y$.
It might be possible that using the standard approach of including static 
3NF one might be able to explain 
the data at higher energies, particularly using Effective 
Field Theory generated interactions \cite{epelbaum:00a,machleidt:11a}. 
Extension of the method to other reactions in the four-nucleon system 
is in progress.


\end{document}